\documentclass{emulateapj}

\begin{document}
\title{Correlation between X-ray Lightcurve Shape and Radio Arrival Time in 
the Vela Pulsar}
\author{A. Lommen\altaffilmark{1}, J. Donovan\altaffilmark{1,2},  
C. Gwinn\altaffilmark{3},
Z. Arzoumanian\altaffilmark{4,5}, A. Harding\altaffilmark{4}, 
M. Strickman\altaffilmark{6}, R. Dodson\altaffilmark{7},
P. McCulloch\altaffilmark{8}, and D. Moffett\altaffilmark{9}}
\altaffiltext{1}{Department of Physics and Astronomy, Franklin and Marshall College, Lancaster, Pennsylvania.}
\altaffiltext{2}{Department of Astronomy, Columbia University, New York, New York.}
\altaffiltext{3}{Department of Physics, University of California, Santa Barbara, California.}
\altaffiltext{4}{NASA Goddard Space Flight Center, Greenbelt, Maryland.}
\altaffiltext{5}{Universities Space Research Association, Columbia, Maryland.}
\altaffiltext{6}{Code 7651.2, Naval Research Laboratory, Washington, DC.}
\altaffiltext{7}{Observatorio Astron\'omico Nacional,
                 Madrid, Espa\~na}
\altaffiltext{8}{University of Tasmania, Tasmania, Australia.}
\altaffiltext{9}{Furman University, Greenville, South Carolina.}

\submitted{Submitted to the Astrophysical Journal, 13 Sep 2006;  Revised, 
	04 Nov 2006; Accepted, 11 Nov 2006}

\begin{abstract}

We report the results of simultaneous observations of the Vela pulsar
in X-rays and radio from the RXTE satellite and the Mount Pleasant
Radio Observatory in Tasmania.  We sought correlations between the
Vela's X-ray emission and radio arrival times on a pulse by pulse
basis. At a confidence level of 99.8\% we have found significantly
higher flux density in Vela's main X-ray peak during radio pulses that
arrived early.  This excess flux shifts to the `trough' following the
2nd X-ray peak during radio pulses that arrive later.  Our results
suggest that the mechanism producing the radio pulses is intimately
connected to the mechanism producing X-rays.  Current models using
resonant absorption of radio emission in the outer magnetosphere as a
cause of the X-ray emission are explored as a possible explanation for
the correlation.
\end{abstract}


\section{Introduction}

The Vela pulsar (PSR B0833$-$45) has been well-studied.  Much
observational work has been done to understand Vela's emission in
individual wavelength regions; e.g. the work of
\nocite{Krishnamohan83} Krishnamohan \& Downs (1983, hereafter KD83)
in the radio regime, observations by \citet{Ogelman93} in the X-ray
regime, and studies by \cite{Kanbach94} in the gamma ray regime.
Observations of Vela's spectrum allow for the possibility of both
polar cap \citep{Daugherty96} and outer-gap \citep{Cheng00} models of
emission.  Vela's pulse profiles in individual regions have been
phase-aligned using the phase of the radio pulse across the optical,
X-ray, and gamma ray wavelength bands \nocite{Harding02} (Harding et
al. 2002, hereafter H02).  This article works to further relate Vela's
high energy emission to its low-energy (radio) emission.

X-ray observations of Vela are challenging.  Though Vela is the
strongest gamma ray source in the sky, the pulsar's spectral power
drops off in the hard X-ray band, making its X-ray emission very
difficult to detect.  Additionally, the pulsed emission is overwhelmed
by the bright but unpulsed background of the X-ray emission nebula in
which the pulsar is embedded \citep{Helfand01}.

The single-peaked pulse profile of the Vela pulsar in radio
wavelengths is much simpler than its high energy counterparts,
although several studies have revealed compound emission. KD83
detected peak-intensity dependent changes in the pulse-shape with the
strongest pulses arriving earlier than the averaged profile.  They
conclude that the radio peak is composed of four different components
originating at different heights in the emission cone with components
lower in the magnetosphere arriving later.  In this article one aspect
we explore is whether a similar connection persists in the X-ray
regime, i.e. whether the X-ray pulses associated with early-arriving
radio pulses have a different shape and/or a different flux than
others.

Related work on other pulsars includes experiments probing the
relationship between the Crab pulsar's ``giant" pulsed emission and
its gamma ray emission (Lundgren et al., 1995, hereafter
Lu95\nocite{Lundgren95}) or its optical emission (Shearer et al.,
2003, hereafter Sh03\nocite{Shearer03}).  They reached opposite
conclusions.  Lu95 observe no correlation within their sensitivity,
indicating that variations in radio flux are caused by changes in
radio coherence, which only affects the radio intensity. Sh03 observe
a significant correlation, and they thus conclude that the increased
emission in the optical and radio is caused by an increased pair
production efficiency.

Patt et al. (1999)\nocite{Patt99} study pulse-to-pulse variability in
the X-ray regime for the Crab pulsar, and they find the pulses to be
steady to 7\%. Using this result, as well as previous work showing
that the Crab exhibits giant radio pulses roughly every two minutes,
they conclude that the radio and X-ray emission mechanisms are not
closely related, even though it is likely that the optical and X-ray
emission regions exist in the same section of the magnetosphere
\citep{Patt99}.

Additional experiments linking pulsar emission in the radio and X-ray
regimes have been performed by Cusumano et
al. (2003)\nocite{Cusumano03} and Vivekanand
(2001)\nocite{Vivekanand01}. Cusumano et al. show that in PSR B1937+21
there is close phase alignment between X-ray pulses and giant radio
pulses, suggesting a correlation in their emission regions.
Vivekanand, on the other hand, finds that the X-ray flux variations
are so much smaller than those at radio wavelengths that they are
inconsistent with the existence of any relationship between the
charged emitters in the two wavelength regimes.

Giant radio pulses have not been shown to exist in Vela, but Johnston
et al. (2001, hereafter J01) \nocite{Johnston01} discovered
microstructure and `giant micropulses' in the Vela pulsar.  The giant
micropulses have flux densities no more than ten times the mean flux
density and have a typical pulse width of $\sim 400 \mu s$.

By doing a pulse-by-pulse analysis of the Vela pulsar in X-ray and
radio wavelengths, we will show in this paper that the Vela pulsar's
X-ray and radio emission mechanisms must be related.  We will discuss
the X-ray and radio observations in
\S \ref{sec.observations}, our analysis in 
\S \ref{sec.analysis}, the effects of scintillation in 
\S \ref{sec.scintillation}, a discussion of interpretations in 
\S \ref{sec.discussion}, and finally our conclusions and related future work in 
\S \ref{sec.conclusion}.

\section{Observations}
\label{sec.observations}

Our data consist of 74 hours of simultaneous radio and X-ray
observations taken over three months at the Mount Pleasant Radio
Observatory in Tasmania and with the RXTE satellite. The radio data
were acquired during 12 separate observations between 30 April and 23
August, 1998.

The radio data were collected as part of the long term monitoring
program of ten young pulsars, including Vela \citep{Lewis03}.  These
data were collected with the 26m antenna at 990.025 MHz using the
incoherently de-dispersed single pulse system (full description in
Dodson, McCulloch, \& Lewis, 2002\nocite{Dodson02}). All individual
pulses from Vela are detectable, and the pulse height, integrated
area, and central time of arrival (for the solar-system barycenter)
were calculated from cross-correlation with a high signal to noise
template in the usual fashion.

The X-ray data were taken during the same three months, yielding 265
ks of usable simultaneous observation. For the purposes of this
project, only top-layer data from RXTE's Proportional Counter Units
(PCUs) in Good Xenon mode in the energy range of 2-16 keV were
used. Other filtering parameters included were standard RXTE criteria:
elevation was greater than 10 degrees, offset was less than 0.02
degrees, the data were taken with at least 3 PCUs on, time since SAA
was at least 30 minutes, and electron0 was less than 0.105.

\section{Analysis} 
\label{sec.analysis}

We filtered the X-ray photon arrival times and transformed them to the
Solar System Barycenter (SSB) using the standard FTOOLS
\citep{Blackburn95} package.  We calculated the pulsar phase at the
time of each X-ray photon, using the radio pulsar-timing program
TEMPO$^{10}$, and the ephemeris downloadable from Princeton
University\footnote{See http://pulsar.princeton.edu/tempo}.  We
matched each X-ray photon with the radio pulse that arrived at the SSB
at the same time.  The precise time span associated with each radio
pulse was given by our best model for arrival time of the peak of the
radio pulse, $\pm 0.5\times$ the instantaneous pulse period calculated
via the model.  Photons arriving on the borderline were associated
with the earlier pulse.  We then compared pulse profiles for X-rays
segregated according to the arrival time of the radio pulse.

\begin{figure}
\plotone{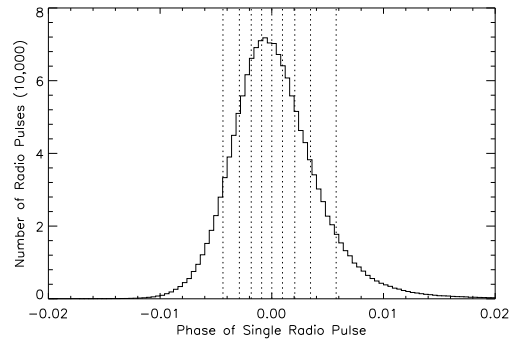}
\caption{The number of radio pulses vs phase relative to a predictive
long-term timing model over the entire data set.  The dotted lines
show the average position of boundaries of the 10 bins that were used
to make the 10 profiles shown in Figure \ref{fig.hists}.
\label{fig.phasedist}
}
\end{figure}

Single radio pulses arrive at a range of times around the predicted
arrival time, as KD83 found.  The histogram of residual arrival times
for radio pulses, relative to the prediction of our best model, is
shown in Figure \ref{fig.phasedist}.  In the figure, and in our
analysis, the average residual from each 5-minute segment of data was
subtracted from all the data in that segment in order to account for
any systematic wandering of the pulse arrival times relative to our
model.  The distribution of arrival times is slightly skewed, with a
tail at late arrival times, so that the mean of the distribution is
slightly later than the mode (at the peak of the histogram), as Figure
\ref{fig.phasedist} shows.  We divided all of the pulses into 10
deciles, by the residual phase of the radio pulse, with equal numbers
of pulses in each decile.  Figure \ref{fig.phasedist} shows our
division of the residual phase of the radio pulse into the deciles.
The deciles are well mixed in time, i.e. no particular observing epoch
dominates any decile.  Removing the 5-minute average residual
eliminates effects of long-term trends that might appear independently
in the radio pulse-timing and X-ray photon counting data.

\begin{figure*}
\plotone{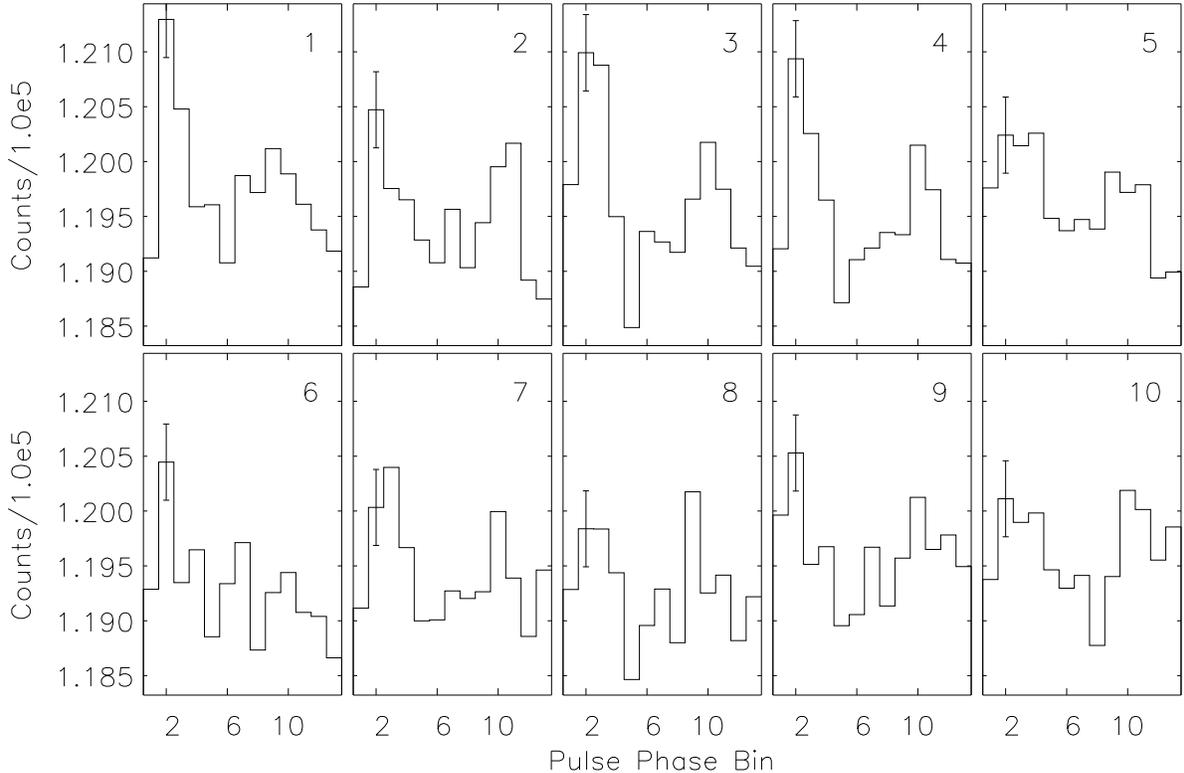}
\caption{ Full-period X-ray lightcurves for photons detected during
radio pulse arrival times falling in the 10 decile bins shown in
Figure \ref{fig.phasedist}.
\label{fig.hists}
}
\end{figure*}

We formed an X-ray pulse profile, integrated from 2$-$16 keV, for each
of the deciles of radio-pulse arrival times, from the X-ray photons
associated with each.  Figure \ref{fig.hists} shows the 10 resulting
X-ray profiles.  Each profile contains 13 bins in phase with the radio
peak being at the left edge of each plot on the border between bins 13
and 1.  The X-ray profiles are significantly different; the X-ray
pulse changes in shape with the arrival of the radio pulse.  We denote
the ten X-ray profiles by their ``lateness": profile 1 comprises the
decile of X-ray photons from the earliest radio-pulse arrivals, and
profile 10 the decile from the latest radio-pulse arrivals.  In
particular it appears that the first X-ray pulse is sharper and
stronger in the earlier deciles.  Overall 2 distinct X-ray peaks are
visible in each of the first 5 profiles, whereas in profiles 6-10 the
two peaks are difficult to distinguish.  In order to quantify these
changes, we performed a number of statistical tests on these 10
profiles including a full Monte Carlo simulation described at the end
of this section.

\begin{figure}[b]
\plotone{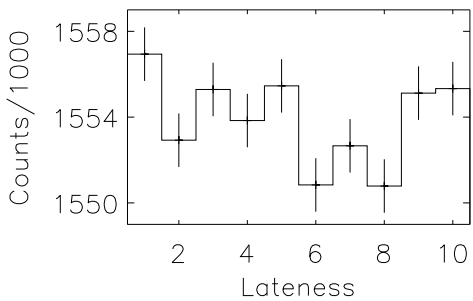}
\caption{
Total X-ray counts vs `lateness' decile.
\label{fig:total_counts}
}
\end{figure}

We observe no significant trend in total X-ray flux with
lateness. Figure \ref{fig:total_counts} shows total counts for each of
the 10 deciles.  From these data, we determine Pearson's correlation
coefficient of $r=-.26$ of total X-ray counts with radio-pulse
lateness with a confidence interval of 54\%.  Nominally, this is not
distinguishable from the hypothesis of zero correlation.  Note,
however, that Pearson's $r$ is not necessarily the best statistic for
this comparison, as we discuss below.

We do find that the total counts reported in Figure
\ref{fig:total_counts} are inconsistent with the Gaussian distribution
expected for this limit of the Poisson distribution produced by shot
noise, at 99.7\% confidence, as determined from a $\chi^2$ test.  This
suggests that there is indeed an evolution of X-ray profile with radio
lateness.  If we consider the counts in individual bins, we find that
they do deviate from the mean more than one would expect from Poission
statistics, as well.  A simple $\chi^2$ test determines that in bins 2
and 3 we can reject the hypothesis of Poisson noise around the mean
value with 81\% and 83\% confidence respectively.  In other words,
bins 2 and 3 show larger changes than we would expect a priori.  The
changes among the 10 profiles in each of the other 11 lateness bins
are consistent with Poisson noise.

The $\chi^2$ test, however, is not suited to detecting trends.  In
order to detect trends we look to Pearson's correlation coefficient,
$r$ and its associated confidence interval.  Pearson's $r$ has limited
validity in our case because it usually assumes that both variables
are drawn from random distributions with nearly Gaussian statistics.
In our case, lateness is deterministic rather than random and
Gaussian.  Shot noise in the profile is random and Gaussian, but
differences of the X-ray profile among deciles need not be.
Nonetheless, this is a useful first step to take to estimate the
significance of any trends.  Note also that Pearson's $r$ itself,
though widely used to measure the strength of a correlation, does not
judge the existence of a correlation.  We therefore place more stake
in the confidence interval, although this involves additional
assumptions about the variables correlated \citep{NR14.5}.  If our
visual impression of the profiles is accurate we would expect that in
the trough, made up of bins 12, 13 and 1, the correlations are
positive (increasing counts with increasing lateness).  All 3 bins do
indeed show positive correlations, 0.29, 0.66 and 0.35 respectively
with associated confidence intervals of 58\%, 96\% and 68\%.  In other
words, 2 of the 3 bins show a significant (greater than 1$\sigma =
65$\%) correlation.  Likewise we would expect that the first X-ray
peak, made up of bins 2 and 3, would show negative correlations
(decreasing counts with increasing lateness).  Bins 2 and 3 do in fact
show negative correlations of $-0.73$ and $-0.48$ with 98\% and 85\%
confidence, consistent with our interpretation of the first peak
decreasing with lateness.  The only other bin that displays a
significant correlation is bin 8, which shows a negative correlation
of $-0.63$ with 95\% confidence.

\begin{figure*}
\plottwo{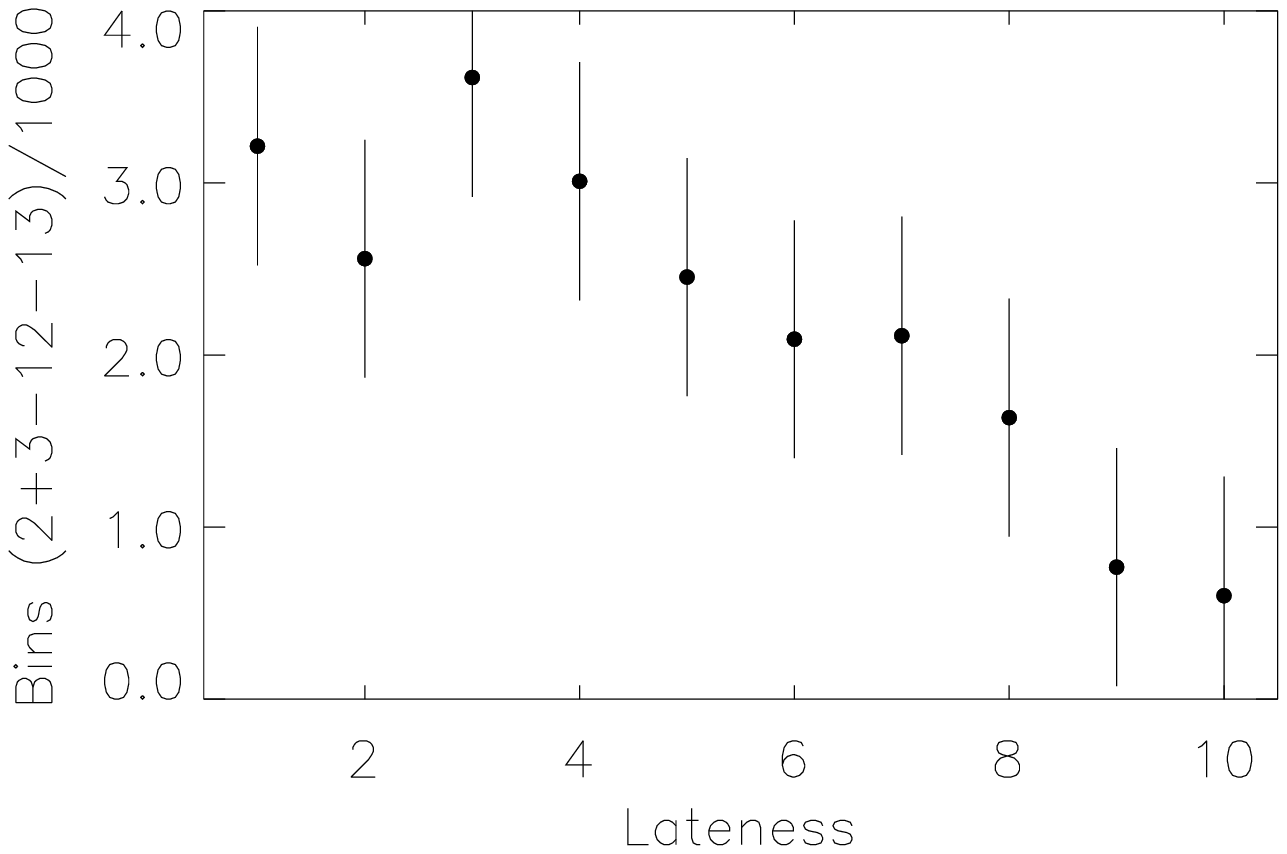}{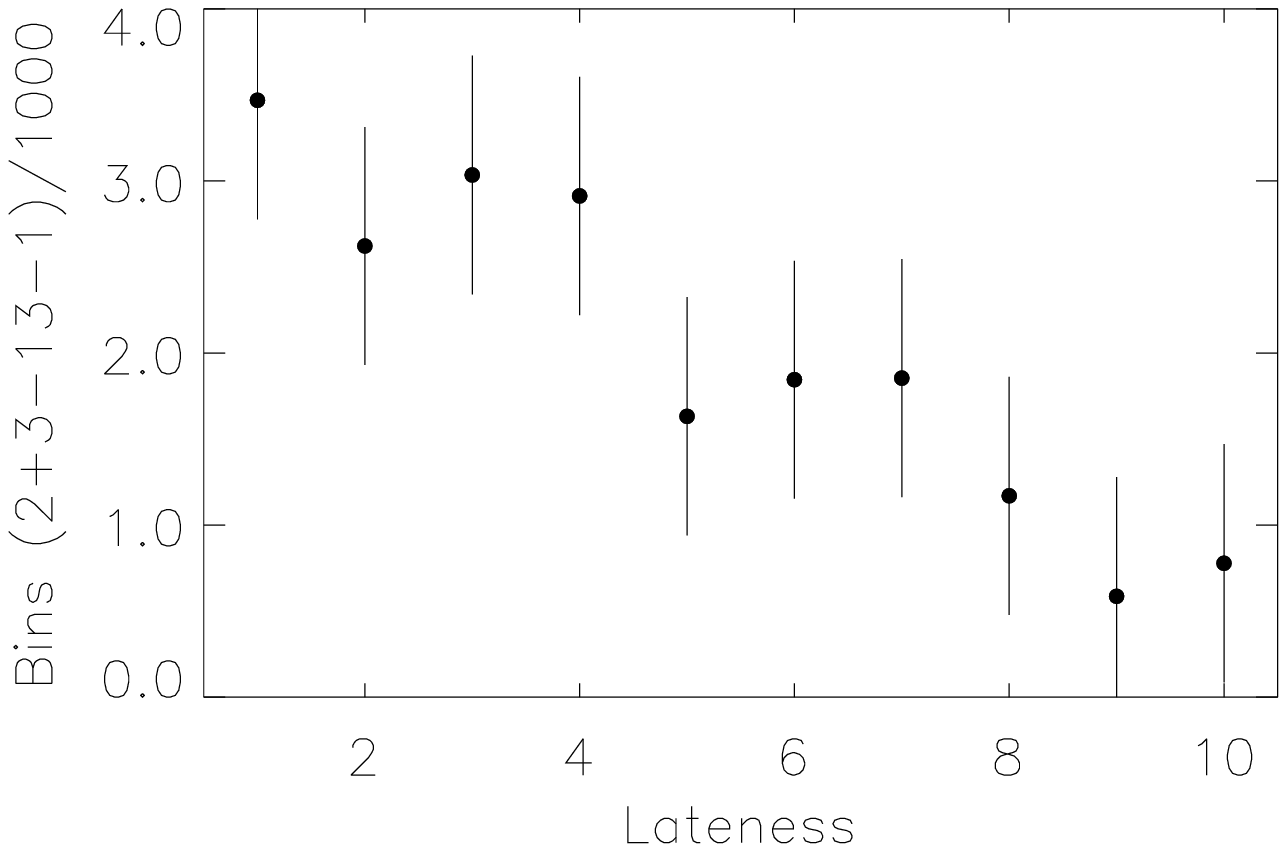}
\caption{
Left:  $Y_a = $bin2 + bin3 $-$ bin12 $-$ bin13 vs the lateness of the radio
pulse.
Right:  $Y_b = $bin2 + bin3 $-$ bin13 $-$ bin1 vs the lateness of the radio
pulse.
\label{fig:plot_compare_10} 
}
\end{figure*}

Next we looked to see if pairs of bins could be combined together to
better detect the signal.  Based on our results described above we
were curious about whether the quantity $Y_{i,a} =
(c_{i\,2}+c_{i\,3})-(c_{i\,12}+c_{i\,13})$ or $Y_{i,b} =
(c_{i\,2}+c_{i\,3})-(c_{i\,13}+c_{i\,1})$ would show significant
trends with lateness.  Here $c_{i\,1}, c_{i\,2}, c_{i\,3}, c_{i\,12},
c_{i\,13}$ are the measured counts in bins 1, 2, 3, 12, and 13, at
lateness $i$.  $Y_{i,a}$ and $Y_{i,b}$ effectively measure, for the 10
profiles, the height of the peak minus the trough and are shown in
Figure \ref{fig:plot_compare_10}.  Via a simple $\chi^2$ test we can
reject with 97\% confidence the hypothesis that the parent
distribution of either of the $Y$'s is a constant at the mean value.
More importantly Figure \ref{fig:plot_compare_10} shows a clear
systematic trend of $Y$ with lateness.  A line fitted to the data and
its associated Poisson uncertainty shown in Figure
\ref{fig:plot_compare_10} yields a slope of $m=-297$ and $-308$
respectively with a formal 1-standard deviation uncertainty of
$\sigma=76$ for both, so both represent (not independently) $\sim 4 $
standard deviation detections of a trend in these data.  For the sake
of completeness we tried all possible differences of pairs of adjacent
bins.  The next highest magnitude slope was 208, but this and all other
significant correlations included some subset of bins 12, 13, 1, 2,
and 3.  The correlation coefficients and associated confidence
intervals for $Y_{i,a}$ and $Y_{i,b}$ are $-0.91$ at 99.97\% and
$-0.84$ at 99.74\%.

Given the limited validity of the Pearson's $r$ coefficient with
non-random, non-normally distributed data, we sought a more rigorous
estimation of the possibility that such a significant trend would
arise amongst 10 such profiles merely by chance, i.e. merely by random
fluctuation of each bin around its mean.  To answer this question we
performed a Monte Carlo simulation with $10^5$ realizations of the
data.  Each simulated data set consisted of 10 profiles, each of 13
bins.  The counts in each bin were chosen as a Gaussian deviate of the
X-ray pulse profile, averaged over lateness, and with variance set by
Poisson statistics.  Thus, in the simulated data sets all differences
among the 10 profiles arose entirely from counting statistics.  In
each simulated data set, for every possible pair of summed adjacent
bins (\{1,2 \& 3,4\}, \{1,2 \& 4,5\} ...\{j,k \& n,o\}, with bins
\{j,k\} adjacent and bins \{n,o\} adjacent: 66 pairs in all) we
computed $Y_i$:
$$
	Y_i = c_{ij}+c_{ik} - (c_{in} + c_{io})
$$ 
where $i=1$ corresponds to the earliest profile in terms of radio
phase, and $c_{ij}$ is the number of counts in the jth bin of the ith
profile.  For the set of points \{$i$, $Y_i$\} where $m$ and $\sigma$
are the slope and uncertainty of the best-fit straight line we
computed the following statistic:
$$
	\Gamma = (m/\sigma)^2
$$ 
As for the fit to our data described above, we weighted the data by
the uncertainties as given by Poisson statistics.  We then compared
the largest $\Gamma$ for each simulated data set to the largest
$\Gamma$ for the actual data (16.4) and found a probability of 0.0024
that the correlation we observe could have been obtained by chance.

We tried other statistics to see if we could find one more
sensitive to the presence of a correlation like that we observed.
The most sensitive one included Pearson's 
$r$ as follows:
$$
	\Gamma^\prime = r^2 + (m/\sigma^\prime)^2
$$ 
where $\sigma^\prime$ is the $\sqrt{{\rm mean}(c_{ij}+c_{ik},
c_{il}+c_{im})}$.  $\Gamma^\prime$ yielded a significance of 0.0007.
$\Gamma^\prime$ is more sensitive than $\Gamma$ to the proximity of
our data to the fitted line, so it yields a smaller probability of
false detection.  Regardless, we retain the more straightforward
$\Gamma$ as a conservative estimate of the significance of our result.

\begin{figure}[b]
\plotone{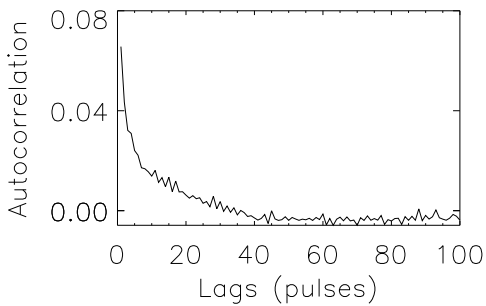}
\caption{
Autocorrelation of radio residual vs lags (pulses).
\label{fig:autocorr} 
}
\end{figure}

Of use in interpreting our results is knowledge of the character of
the radio residuals by which we are binning the X-ray photons.  KD83
did much work in this area, but one question they do not address
directly is the following: if a particular pulse is ``early", what is
the chance that the next pulse will also be early?  We calculated the
autocorrelation function of radio residual, shown in figure
\ref{fig:autocorr}.  The function is normalized by the autocorrelation
at zero lag.  The function at a lag of 1 pulse is represented by the
left edge of the plotted curve, at a value of 0.066.  The figure shows
that the pulsar has very little ``memory" of the lateness of the
previous pulse, although it is interesting to note there is finite
correlation out to 40 pulse periods.

\section{Scintillation}
\label{sec.scintillation}

In contrast to effects intrinsic to the pulsar, scintillation is
unlikely to produce the observed association, because it does not
affect X-rays; scintillation might erase such a correlation but it
cannot introduce it. Diffractive scintillation for Vela at our
frequency has a characteristic bandwidth of 2 kHz (Gupta 1995), far
smaller than our observing bandwidth of 6.4 MHz. We therefore average
over $\sim$3200 independent scintillation elements. Refractive
scintillation modulates flux density over a wide bandwidth and has
timescale $\sim$25 days \citep{Desai92}.  This was shorter than the
total span of the observations, but much longer than the span of any
one observation.  To combat any effect that this might have on the
observed arrival times, we defined the 10 deciles separately for each
observation; i.e. the specific values of residual that separated each
of the 10 bins were calculated for each individual radio
observation. Again, these cutoffs were defined in such a way that an
equal number of radio pulses was associated with each decile.

We assumed that the pulsar was stationary relative to a refractive
scintillation element during each single observation, and further
tests to ensure that the length of the observation was not a factor
were performed by renormalizing the cutoffs using both one hour and
five minute timespans. Reanalyzing the X-ray data in one hour segments
did not significantly change our results, and the five minute spans
were found to be too short to accurately represent Vela's emission.

\section{Discussion}
\label{sec.discussion}

The early radio emission could result from coherent radiation along a
different set of field lines (i.e. more leading) or from radiation at
a higher altitude, or both. The work of KD83 suggests both. The early
radio emission may also result from stochastic fluctuations in the
radio beam intensity which would lead to pulse arrival time changes by
changing the shape of the radio beam as it takes a finite time to
sweep across our line of sight.

Petrova \nocite{Petrova03} (2003, and references therein) offers a
physical model that could explain the observed relationship between
the radio and X-ray emission.  Her model suggests that resonant
absorption of radio emission from the outer magnetosphere leads to an
increase in the pitch angles and momenta of the secondary pairs, which
then leads to optical and higher energy emission by spontaneous
synchrotron radiation.  How could this model produce a changing
X-ray pulse shape?  Due to the effects of rotation of the
magnetosphere and aberration, the early radio emission can cross a
larger number of higher altitude field lines and at larger angles,
thereby maximizing the opportunity for absorption by particles on
those field lines, and therefore the production of high-energy
radiation. Conversely a late radio photon on a path almost directly
along the magnetic pole may escape the magnetosphere with many fewer
interactions, since it will cross fewer open field lines and at
smaller angle.  The details of the above are somewhat unimportant as
our knowledge of the magnetosphere and the plasma therein is limited,
but the point is that as different parts of the radio beam are active,
resonant absorption may happen at different rates in different parts
of the magnetosphere, causing continuous change in the shape of the
observed X-ray emission.

More generally our results imply a connection between the radio and
X-ray emission mechanisms for Vela that is not consistent with outer
gap models. In these models, the high energy emission results from a
gap connection to the pole opposite from that producing the radio
emission. The pole and outer gap associated with the same set of field
lines are not visible to one observer.  It is not clear how a
correlation could exist between the radio and high energy regimes in
these models.

The giant micropulse emission observed by J01 would cause a single
radio pulse to arrive about 1 ms early, so it is realistic to consider
the possibility that the giant micropulse emission is primarily
responsible for the early arrival of the radio pulse.  However, out of
20,085 pulses that J01 observed at 1413 MHz, 14 of them contained
giant micropulses.  So the giant micropulses may be influencing the
first of our 10 deciles, but cannot contribute to the effect observed
in the other 9 deciles.  We conclude that Vela's giant micropulse
emission cannot be responsible for the effect presented here.

\section{Conclusions and Future Work}
\label{sec.conclusion}

We have evidence that links features of Vela's X-ray emission with
features of its radio emission.  First, we find that X-ray pulses
associated with early radio pulses show stronger emission at the main
X-ray peak which is the sharper of the two.  Similarly X-ray pulses
associated with later radio pulses show stronger emission at the
trough following the 2nd X-ray peak, a region in phase near the radio
peak.  The trend we measure has a 0.2\% probability of appearing in
the data by chance.  We conclude that there is a close relationship
between X-ray and radio emission in the Vela pulsar.

We plan to further characterize the relationship between the radio and
high energy emissions of pulsars to identify their origins and
constrain magnetospheric models.  In particular, we will explore the
dependence of radio-to-X-ray correlations on the radio frequency and
polarization properties of individual Vela pulses, both of which carry
information about emission altitudes. Similar observations of other
pulsars also promise useful insights as probes of different magnetic
field strengths and emission/viewing geometries.

\acknowledgments

Many thanks to Ben Stappers, Michael Kramer, Russell Edwards, Wim
Hermsen, David Helfand, Paul Ray, Simone Migliari and Tiziana Di Salvo
for helpful comments.  AL is grateful for the hospitality of the
Australia Telescope National Facility and for a Research Corporation
Grant in support of this research.  ZA was supported by NASA grant
NRA-99-01-LTSA-070.  CG acknowledges support of NSF-AST-9731584.  AH
acknowledges support from the NASA Astrophysics Theory Program.  RD
acknowledges support as a Marie-Curie fellow via EU FP6 grant
MIF1-CT-2005-021873.  This research has made use of data obtained from
the High Energy Astrophysics Science Archive Research Center
(HEASARC), provided by NASA's Goddard Space Flight Center.


\begin{thebibliography}{}

\bibitem[{Blackburn}(1995){Blackburn}]{Blackburn95}
{Blackburn}, J.~K. 1995, In ASP Conf. Ser. 77: Astronomical Data Analysis
  Software and Systems IV, R.~A. {Shaw}, H.~E. {Payne}, and J.~J.~E. {Hayes},
  eds., p. 367

\bibitem[{Cheng}, {Ruderman}, \& {Zhang}(2000){Cheng}, {Ruderman}, and
  {Zhang}]{Cheng00}
{Cheng}, K.~S., {Ruderman}, M., \& {Zhang}, L. 2000, \apj, 537, 964

\bibitem[{Cusumano} {et~al.}(2003){Cusumano}, {Hermsen}, {Kramer}, {Kuiper},
  {L{\"o}hmer}, {Massaro}, {Mineo}, {Nicastro}, and {Stappers}]{Cusumano03}
{Cusumano}, G., {Hermsen}, W., {Kramer}, M., {Kuiper}, L., {L{\"o}hmer}, O.,
  {Massaro}, E., {Mineo}, T., {Nicastro}, L., \& {Stappers}, B.~W. 2003, \aap,
  410, L9

\bibitem[{Daugherty} \& {Harding}(1996){Daugherty} and {Harding}]{Daugherty96}
{Daugherty}, J.~K., \& {Harding}, A.~K. 1996, \apj, 458, 278

\bibitem[{Desai} {et~al.}(1992){Desai}, {Gwinn}, {Reynolds}, {King}, {Jauncey},
  {Flanagan}, {Nicolson}, {Preston}, and {Jones}]{Desai92}
{Desai}, K.~M., {Gwinn}, C.~R., {Reynolds}, J., {King}, E.~A., {Jauncey}, D.,
  {Flanagan}, C., {Nicolson}, G., {Preston}, R.~A., \& {Jones}, D.~L. 1992,
  \apjl, 393, L75

\bibitem[{Dodson}, {McCulloch}, \& {Lewis}(2002){Dodson}, {McCulloch}, and
  {Lewis}]{Dodson02}
{Dodson}, R.~G., {McCulloch}, P.~M., \& {Lewis}, D.~R. 2002, \apjl, 564, L85

\bibitem[{Harding} {et~al.}(2002){Harding}, {Strickman}, {Gwinn}, {Dodson},
  {Moffet}, and {McCulloch}]{Harding02}
{Harding}, A.~K., {Strickman}, M.~S., {Gwinn}, C., {Dodson}, R., {Moffet}, D.,
  \& {McCulloch}, P. 2002, \apj, 576, 376

\bibitem[{Helfand}, {Gotthelf}, \& {Halpern}(2001){Helfand}, {Gotthelf}, and
  {Halpern}]{Helfand01}
{Helfand}, D.~J., {Gotthelf}, E.~V., \& {Halpern}, J.~P. 2001, \apj, 556, 380

\bibitem[{Johnston} {et~al.}(2001){Johnston}, {van Straten}, {Kramer}, and
  {Bailes}]{Johnston01}
{Johnston}, S., {van Straten}, W., {Kramer}, M., \& {Bailes}, M. 2001, \apjl,
  549, L101

\bibitem[{Kanbach} {et~al.}(1994){Kanbach}, {Arzoumanian}, {Bertsch},
  {Brazier}, {Chiang}, {Fichtel}, {Fierro}, {Hartman}, {Hunter}, {Kniffen},
  {Lin}, {Mattox}, {Mayer-Hasselwander}, {Michelson}, {von Montigny}, {Nel},
  {Nice}, {Nolan}, {Pinkau}, {Rothermel}, {Schneid}, {Sommer}, {Sreekumar},
  {Taylor}, and {Thompson}]{Kanbach94}
{Kanbach}, G., {Arzoumanian}, Z., {Bertsch}, D.~L., {Brazier}, K.~T.~S.,
  {Chiang}, J., {Fichtel}, C.~E., {Fierro}, J.~M., {Hartman}, R.~C., {Hunter},
  S.~D., {Kniffen}, D.~A., {Lin}, Y.~C., {Mattox}, J.~R., {Mayer-Hasselwander},
  H.~A., {Michelson}, P.~F., {von Montigny}, C., {Nel}, H.~I., {Nice}, D.,
  {Nolan}, P.~L., {Pinkau}, K., {Rothermel}, H., {Schneid}, E., {Sommer}, M.,
  {Sreekumar}, P., {Taylor}, J.~H., \& {Thompson}, D.~J. 1994, \aap, 289, 855

\bibitem[{Krishnamohan} \& {Downs}(1983){Krishnamohan} and
  {Downs}]{Krishnamohan83}
{Krishnamohan}, S., \& {Downs}, G.~S. 1983, \apj, 265, 372

\bibitem[{Lewis} {et~al.}(2003){Lewis}, {Dodson}, {Ramsdale}, and
  {McCulloch}]{Lewis03}
{Lewis}, D.~R., {Dodson}, R.~G., {Ramsdale}, P.~D., \& {McCulloch}, P.~M. 2003,
  In ASP Conf. Ser. 302: Radio Pulsars, M.~{Bailes}, D.~J. {Nice}, and S.~E.
  {Thorsett}, eds., p. 121

\bibitem[{Lundgren} {et~al.}(1995){Lundgren}, {Cordes}, {Ulmer}, {Matz},
  {Lomatch}, {Foster}, and {Hankins}]{Lundgren95}
{Lundgren}, S.~C., {Cordes}, J.~M., {Ulmer}, M., {Matz}, S.~M., {Lomatch}, S.,
  {Foster}, R.~S., \& {Hankins}, T. 1995, \apj, 453, 433

\bibitem[{Oegelman}, {Finley}, \& {Zimmerman}(1993){Oegelman}, {Finley}, and
  {Zimmerman}]{Ogelman93}
{Oegelman}, H., {Finley}, J.~P., \& {Zimmerman}, H.~U. 1993, \nat, 361, 136

\bibitem[{Patt} {et~al.}(1999){Patt}, {Ulmer}, {Zhang}, {Cordes}, and
  {Arzoumanian}]{Patt99}
{Patt}, B.~L., {Ulmer}, M.~P., {Zhang}, W., {Cordes}, J.~M., \& {Arzoumanian},
  Z. 1999, \apj, 522, 440

\bibitem[{Petrova}(2003){Petrova}]{Petrova03}
{Petrova}, S.~A. 2003, \mnras, 340, 1229

\bibitem[Press {et~al.}(1992)Press, Teukolsky, Vetterling, and
  Flannery]{NR14.5}
Press, W.~H., Teukolsky, S.~A., Vetterling, W.~T., \& Flannery, B.~P. 1992.
\newblock Numerical Recipes in C, Cambridge University Press

\bibitem[{Shearer} {et~al.}(2003){Shearer}, {Stappers}, {O'Connor}, {Golden},
  {Strom}, {Redfern}, and {Ryan}]{Shearer03}
{Shearer}, A., {Stappers}, B., {O'Connor}, P., {Golden}, A., {Strom}, R.,
  {Redfern}, M., \& {Ryan}, O. 2003, Science, 301, 493

\bibitem[{Vivekanand}(2001){Vivekanand}]{Vivekanand01}
{Vivekanand}, M. 2001, \mnras, 326, L33

\end{thebibliography}
\end{document}